\documentclass{ws-procs975x65}
\usepackage{graphicx, wrapfig}
\def\beq{\begin{equation}}
\def\eeq{\end{equation}}
\def\fnl{f_{_{\rm NL}}}

\def\f{\frac}
\def\h{\hat}
\def\v{\vec}

\def\d{\textrm{d}}

\def\pp{p_{\phi}}

\def\dph{\delta\phi}

\def\dphi{\partial_\phi}

\def\Mpl{M_{\rm Pl}}

\def\phib{\phi_{\rm B}}
\def\rhob{\rho_{\rm B}}

\def\l{\left}
\def\r{\right}

\begin{document}

\title{Computation of non-Gaussianity in loop quantum cosmology}
\author{Vijayakumar Sreenath}

\address{Inter-University Centre for Astronomy and Astrophysics, Ganeshkhind,\\
Pune, Maharashtra 411007, India\\
E-mail: vsreenath@iucaa.in}

\author{Ivan Agullo} 

\address{Department of Physics and Astronomy, Louisiana State University,\\
Baton Rouge, LA 70803, U.S.A.\\
E-mail: agullo@lsu.edu}

\author{Boris Bolliet} 

\address{Jodrell Bank Centre for Astrophysics, The University of Manchester, Alan Turing Building, \\
Oxford Road, Manchester, M13 9PL.\\
E-mail: boris.bolliet@manchester.ac.uk}

\begin{abstract}
We summarize our investigations 
of the second-order perturbations in loop quantum cosmology (LQC). We shall discuss, primarily, two aspects. 
Firstly, whether the second-order contributions arising from the cosmic bounce, 
occurring at Planck scale, could be large enough to break the validity of perturbation 
theory. Secondly, the implications of the upper bounds on primordial non-Gaussianity, 
arrived at by the Planck collaboration, on the LQC phenomenology. 
\end{abstract}


\bodymatter


\section{Introduction}
Loop quantum cosmology (LQC) provides an extension of 
the inflationary paradigm to the Planck era (see, for instance, Ref.~\refcite{Agullo:2016tjh}). 
Over the past decade or so, there has been a research program aimed at 
investigating the viability of LQC as a theory of the pre-inflationary universe. 
Until now, investigations of primordial perturbations generated in LQC 
have focused mainly at the level of the power spectrum. 
In this work, we extend the analysis to the level of three-point functions, namely, the 
bispectrum of curvature perturbations. 
We will analyze primarily two aspects. 
Firstly, we check whether next-to-leading order corrections to the power spectrum 
are sub-leading.
Secondly, we verify that the amount of non-Gaussianity as quantified by the dimensionless 
quantity $\fnl$ is compatible with the observations of cosmic microwave background (CMB) 
and investigate new predictions. 
\section{Computation of the Bispectrum in the Dressed Metric Approach}\label{sec:2}
The system of interest is scalar perturbations $\delta\phi$ living on a 
Friedmann-Lemaitre-Robertson-Walker (FLRW) metric sourced by a scalar field $\phi$. 
In LQC, such a system is described by a wavefunction $\Psi(v,\,\phi,\,\delta\phi)$, where 
$v\,\equiv\,a^3\,{\cal V}_0\,4/\kappa$ with $a$ being the scale factor and ${\cal V}_0$, the 
volume of the universe, introduced to regulate infrared divergence. 
The dynamics is governed by the constraint equation, $\hat{\cal H}\,\Psi\,=\,0$, 
where the Hamiltonian operator can be split in to the background and perturbed 
part as $\hat{\cal H}\, =\, \hat{\cal H}_{_{FLRW}}\, +\, \hat{\cal H}_{pert}$. 
We are interested in solutions wherein $\Psi(v,\,\phi,\,\delta\phi)\,=\,\Psi_0(v,\,\phi,)\,\otimes\,\delta\Psi(v,\,\phi,\,\delta\phi)$, 
where $\Psi_0$ describes a quantum FLRW geometry and $\delta\Psi$ describes 
the scalar perturbations.
\par
The states $\Psi_0$ satisfies the equation $\hat{\cal H}_{_{FLRW}}\,\Psi_0\,=\,0$.  
It has been shown that, for states that are sharply peaked in the volume $v$ 
during the entire evolution, the background geometry can be described by an 
effective classical Hamiltonian (see e.g. Ref.~\refcite{Agullo:2016tjh} and references therein). 
In the dressed metric approach, we are interested in quantum states $\delta\Psi(v,\,\phi,\,\delta\phi)$ that are a 
small perturbation around such a quantum FLRW state $\Psi_0(v,\,\phi)$.  
A detailed analysis shows that $\delta\Psi(v,\,\phi,\,\delta\phi)$ 
are solutions to the Schr\"{o}dinger equation, 
$i \hbar \, \dphi \delta \Psi =  \langle \Psi_0| \hat{\mathcal{H}}_{\rm pert}[N_{\phi}]| \Psi_0 \rangle \,  \delta \Psi$, 
where $\hat{\cal H}_{pert}\,=\,\hat{\cal H}^{(2)}\,+\,\hat{\cal H}^{(3)}$, namely 
the Hamiltonian at second and third order in perturbations respectively and 
$N_{\phi}$ is the lapse associated with relational time $\phi$.\cite{Agullo:2017eyh} 
\par
We are interested in computing the correlation functions of these scalar 
perturbations. 
The first step is to expand the perturbations in Fourier space and introduce 
creation and annihilation operators 
\begin{equation}\label{eq:modeexp}
 \h{\dph}({\vec x},\eta) = \int\f{{\rm d}^3 k}{(2\pi)^3} \left(\hat A_{\vec k}~\varphi_k(\eta) + \hat A^\dagger_{-\vec k}
~\varphi_{k}^*(\eta)\right) e^{i{\vec k}\cdot{\vec x}},
\end{equation}
 where $[\hat A_{\vec k},\hat A^{\dagger}_{\vec k'}]=\hbar \, (2\pi)^3\, \delta^{{(3)}}(\v k+\v k')$ 
 and $[\hat A_{\vec k},\hat A_{\vec k'}]\,=\,0$. 
 The dynamics of perturbations are governed by the second-order 
 Hamiltonian with the background quantities determined using the effective background Hamiltonian. 
 The scalar power spectrum of $\hat\dph$ is defined as 
 \begin{equation}\label{eq:dphi2pf}
  \langle 0|\h{\dph}_{\vec k}( \eta) \h{\dph}_{\vec k^\prime}(\eta)|0\rangle \equiv 
(2\pi)^3\delta^{{(3)}}({\vec k}+{\vec k^\prime}) \f{2\pi^2}{k^3} \mathcal P_{\dph}(k, \eta)\, ,
 \end{equation}
where $|0\rangle$ is the vacuum annihilated by the operators $\hat A_{\vec{k}}$ for all $\vec{k}$. 
For the purpose of relating perturbations to the late time physics, it is convenient to 
express the power spectrum in terms of comoving curvature perturbations. 
The power spectrum of curvature perturbation, $\mathcal R$, in terms of inflaton perturbation $\dph$, evaluated at the 
end of inflation is
$
\mathcal{P}_{\mathcal{R}}(k)\equiv 
\bigg(\frac{a(\eta_{\rm end})}{z(\eta_{\rm end})}\bigg)^2\, \f{\hbar\,{k^3}}{{2\pi^2}}\,|\varphi_k( \eta_{\rm end})|^2\, ,
$
where $z= -\f{6}{\kappa}\f{\pp}{\pi_a}$ with $\kappa\, = 8\,\pi\,G$ and $\pp$ and $\pi_a$ are momenta conjugate to $\phi$ and $a$ respectively. 
\par
The self-interaction of scalar perturbation, at lowest order, is described by the 
third-order interaction Hamiltonian, $\h{\mathcal{H}}_{\rm int} \equiv  \langle \Psi_0| \hat{\mathcal{H}}^{(3)}[N_{\phi}]| \Psi_0 \rangle$.\cite{Maldacena:2002vr}
The perturbations at this order are quantified using the scalar bispectrum, $B_{\mathcal{R}}(k_1,k_2,k_3)$, that is defined in terms of curvature perturbations by 
\begin{equation}\label{fnlz}
\langle 0|\h{\mathcal{R}}_{{\vec k}_1} \h{\mathcal{R}}_{{\vec k}_2} \h{\mathcal{R}}_{{\vec k}_3}|0\rangle
\equiv  (2\pi)^3\delta^{(3)}(\v{k}_1+\v{k}_2+\v{k}_3) \, B_{\mathcal{R}}(k_1,k_2,k_3) \, .
\end{equation}
It is often convenient to quantify the bispectrum using a dimensionless function, $\fnl$, 
which can be defined as, 
\begin{equation}\label{eq:fNLdef} 
B_{\mathcal{R}}(k_1,k_2,k_3)\equiv  -\f{6}{5} \, f_{_{\rm NL}}(k_1,k_2,k_3)\, \times (\Delta_{k_1}\Delta_{k_2}+\Delta_{k_1}\Delta_{k_3}+\Delta_{k_2}\Delta_{k_3})\,, 
\end{equation}
where $\Delta_{k}\equiv  \frac{2\pi^2}{k^3} \, \mathcal{P}_{\mathcal{R}}(k)$ is the dimensionful power spectrum. 
\par
In order to compute bispectrum, we need to express it in terms of $\dph$ as follows, 
\begin{eqnarray} \label{eq:3pf}
&&\langle 0|\h{\mathcal{R}}_{{\vec k}_1} \h{\mathcal{R}}_{{\vec k}_2} \h{\mathcal{R}}_{{\vec k}_3}|0\rangle =
\left(-\f{a}{z}\right)^3  \langle 0|\h{\dph}_{{\vec k}_1} \h{\dph}_{{\vec k}_2} \h{\dph}_{{\vec k}_3}|0\rangle
\, + \left(-\f{3}{2}+3\f{V_{\phi}\, a^5}{\kappa\, \pp\, \pi_a}+\f{{\kappa}}{4}\f{z^2}{a^2}\right)\, \left(-\f{a}{z}\right)^4\,\nonumber \\ 
&& \times\Big[\int \f{d^3p}{(2\pi)^3} \, \langle 0|\h{\dph}_{{\vec k}_1} \h{\dph}_{{\vec k}_2}  \h{\dph}_{{\vec p}}\,  
\h{\dph}_{{\vec k}_3-\v p}|0\rangle + (\v k_1 \leftrightarrow \v k_3)+ (\v k_2 \leftrightarrow \v k_3) \,+\cdots \Big]\, ,
\end{eqnarray}
 where the symbols $(\v k_i \leftrightarrow \v k_j)$ indicate terms obtained by replacing $k_i$ with $k_j$ in the
 first term of the second line and the dots indicate higher order terms. 
At leading order in perturbations, the first term on RHS can be evaluated using 
time dependent perturbation theory 
\begin{eqnarray}\label{eq:3pf-1}
 \langle 0|\h{\dph}_{{\vec k}_1}(\eta) \h{\dph}_{{\vec k}_2}(\eta) \h{\dph}_{{\vec k}_3}( \eta)|0\rangle  &=&
 \langle 0| \h{\dph}^{\rm I}_{{\vec k}_1}( \eta) \h{\dph}^{\rm I}_{{\vec k}_2} ( \eta)\h{\dph}^{\rm I}_{{\vec k}_3}( \eta)|0\rangle\nonumber \\
 &-&\,i/\hbar  \int d \eta' \langle 0|\left[ \h{\dph}^{\rm I}_{{\vec k}_1}( \eta) \h{\dph}^{\rm I}_{{\vec k}_2}(\eta) 
 \h{\dph}^{\rm I}_{{\vec k}_3}( \eta), \h{\mathcal H}^{\rm I}_{\rm int}( \eta')\right]|0\rangle\, , \;\;\;\;\;\;\;\;
\end{eqnarray}
where the superscript $I$ indicates fields in the interaction picture. 
Since, $\dph^I$ is a Gaussian field, the first term vanishes and only the second term 
contributes. 
The second term in the RHS of Eq. (\ref{eq:3pf}) can be evaluated using Wick's 
theorem and Eq. (\ref{eq:dphi2pf}). 
Using Eqs. (\ref{eq:3pf}) and (\ref{eq:3pf-1}), one can compute the bispectrum 
and hence the function $\fnl$ using Eq. (\ref{eq:fNLdef}).  
\section{Numerical Method and Results}\label{sec:3}
In this section, we will briefly describe our implementation of the formalism for 
computing $\fnl$ and the results we obtain. 
In order to compute $\fnl$ at the end of inflation, one needs to evolve the 
perturbations from an early time before the bounce until the perturbations 
leave the horizon during inflation at which point their amplitude  freezes in time. 
We need to make three choices to do this computation. 
Firstly, we need to specify the potential governing the field $\phi$. 
We choose the quadratic potential, $V(\phi)\,=\, m^2\,\phi^2/2$, 
where $m=6.4\times10^{-6}M_{P\ell}$.
Secondly, we need to choose a background geometry by 
specifying the value of $\phi$ and energy density, $\rho$, at the bounce.
We work with $\phi_{\rm B}=7.62\,M_{P\ell}$ and $\rho_{B}=1\,M_{P\ell}^4$, 
where subscript $B$ denotes the bounce, so that the effects due to LQC appear at observable 
scales while respecting the Planck constraints on power spectrum. 
Finally, we need to choose an initial state for perturbations, which we choose to be a Minkowski 
initial state. 
More specifically, we choose $\varphi_k(\eta_0)=\f{1}{a(\eta_0)\sqrt{2\, k}}$ and 
$\varphi'_k(\eta_0)=[-i\, k+\frac{a'(\eta_0)}{a(\eta_0)}]\, \varphi_k(\eta_0)$ 
as initial data for the modes, at conformal time $\eta_0=-2.8\times 10^{3}\,T_{P\ell}$ 
(the bounce takes place at $\eta=0$). 
The initial time was chosen so that all the modes of interest, namely 
those between $k_{\rm min}= k_*/10$, and $k_{\rm max}=1000 k_*$, where 
$k_*/a(t_{today})\, =\, 0.002\, {\rm Mpc}^{-1}$ is the pivot scale, were 
in the adiabatic regime.
We have investigated the effects of varying these choices in detail in Ref.~\refcite{Agullo:2017eyh}.
\par
To perform this computation, we use the platform provided by \verb|class|. \cite{2011JCAP...07..034B}
The computation was done in two stages. 
In the first stage we evolve the background from very early times to the end of inflation. 
In the second step, we convert the time integral in Eq. (\ref{eq:3pf-1}) to a 
differential equation and evolve it together with the differential equation for 
the fourier modes. 
In the remaining part of this section, we will discuss the various results. 
\begin{figure}
\begin{center}
 \begin{tabular}{cc}
  \includegraphics[width=0.45\linewidth]{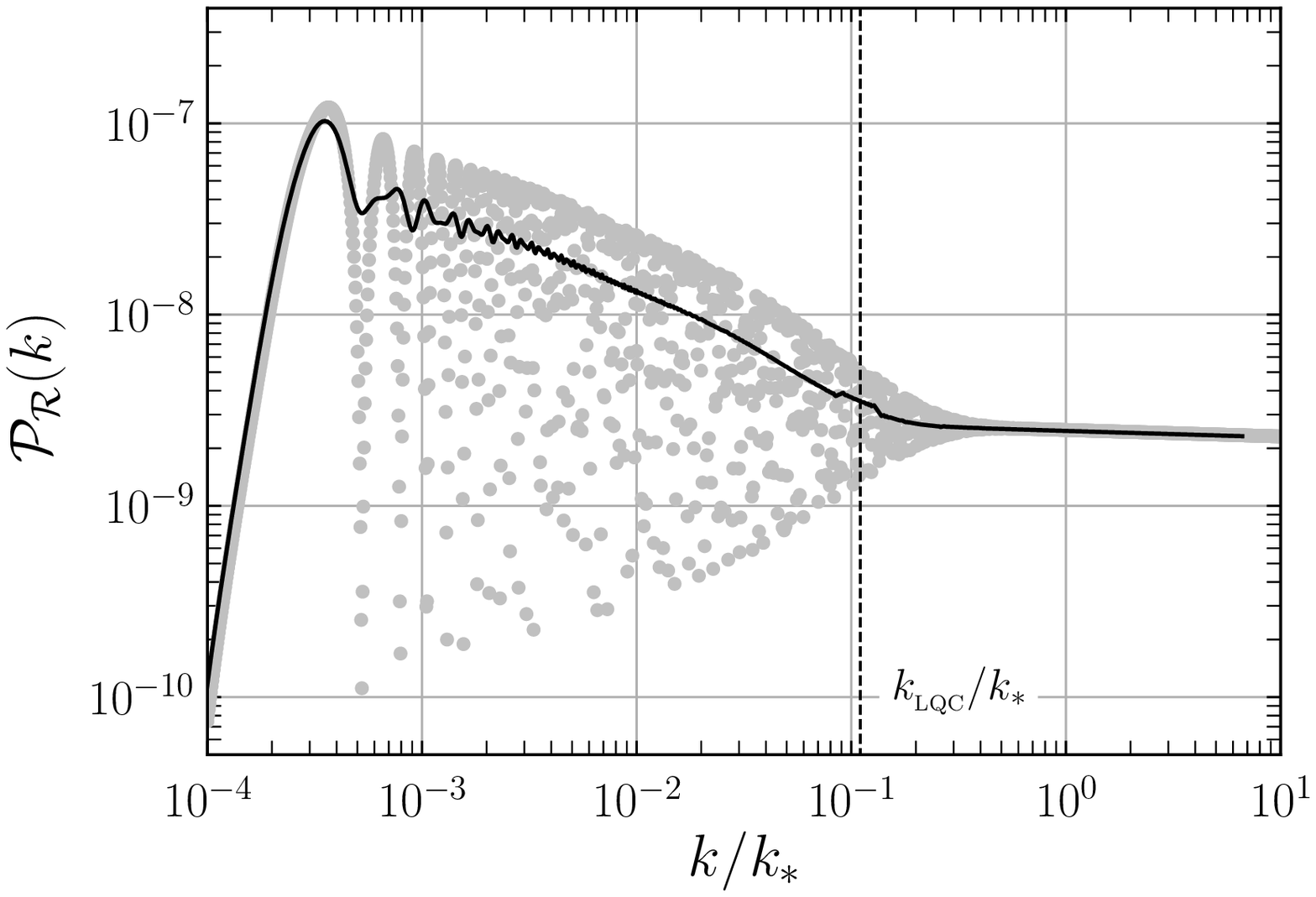} &
  \includegraphics[width=0.45\linewidth]{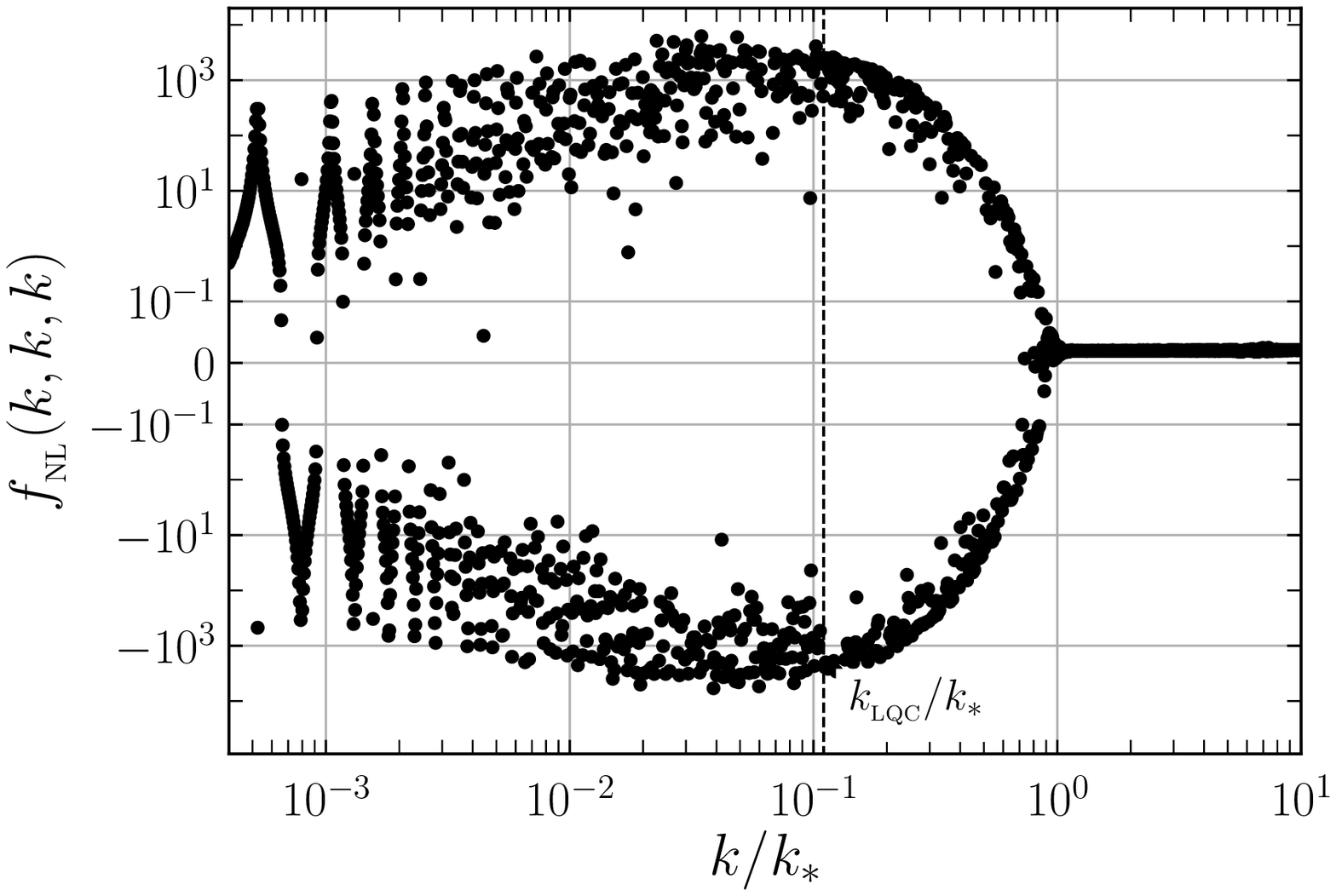}
 \end{tabular}
 \caption{\label{fig:1}The power spectrum and $\fnl(k_1,\,k_2,\,k_3)$ evaluated in the equilateral limit. 
 }
 \end{center}
 \vskip -10pt
\end{figure} 
\begin{figure}
\begin{center}
\begin{tabular}{cc}
  \includegraphics[width=0.45\linewidth]{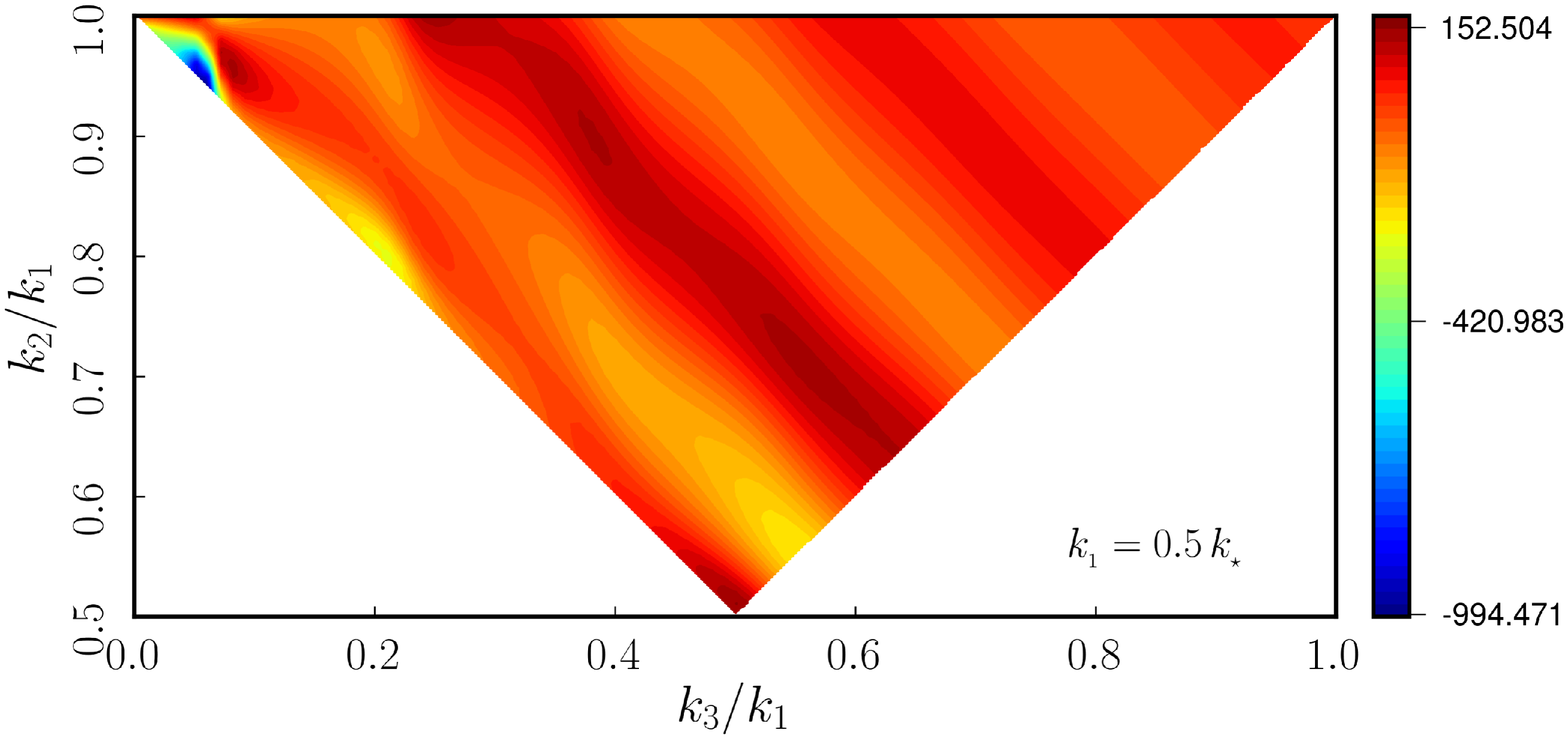} &
  \includegraphics[width=0.45\linewidth]{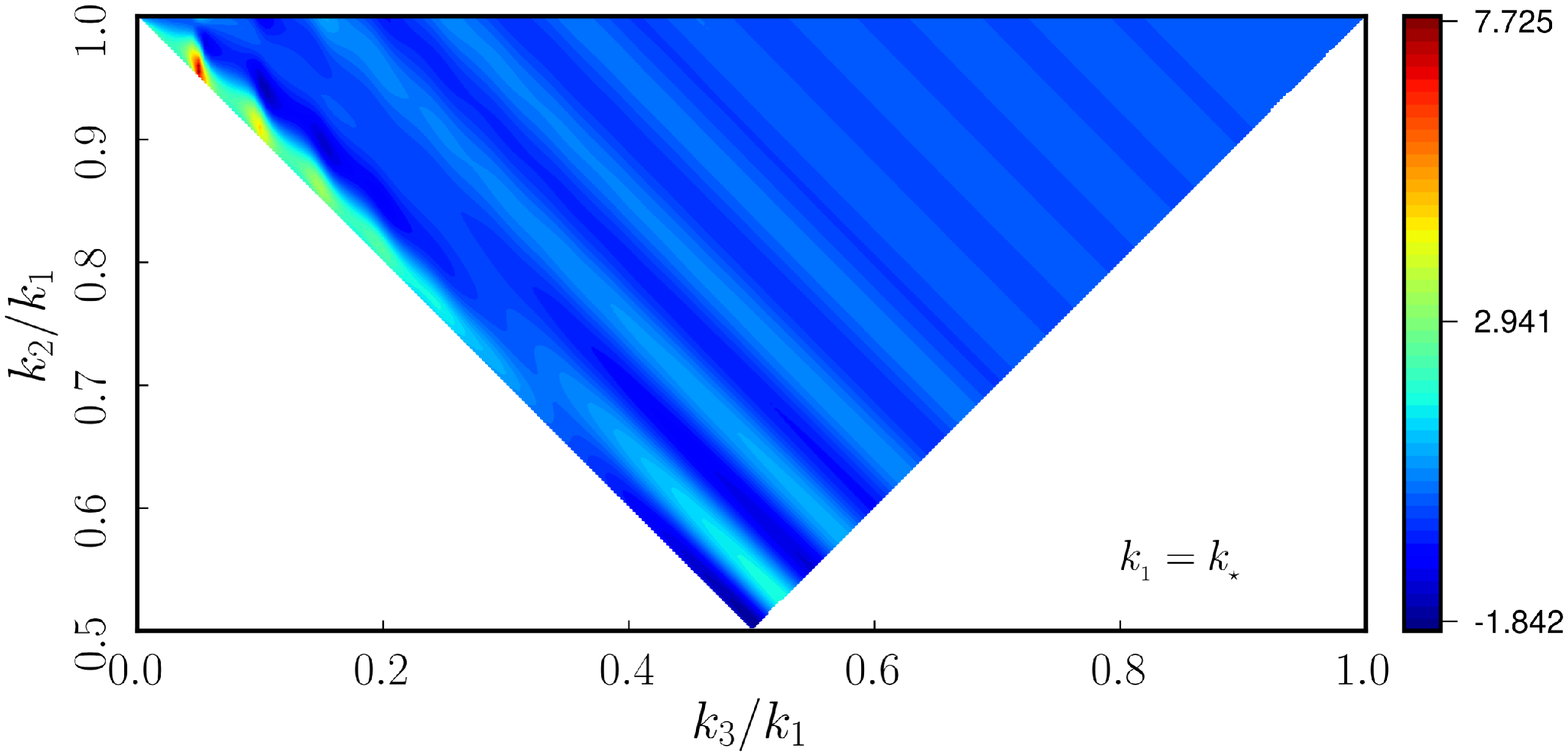}
  \end{tabular}
 \caption{\label{fig:2}The shape of the non-Gaussianity, $\fnl(k_1,\,k_2,\,k_3)$, evaluated  with $k_1\,=\,k_*/2$ and $k_*$.  
 The figure shows the portions allowed by the condition $\v k_1\, +\, \v k_2\, +\, \v k_3\, =\, 0$. 
}
 \end{center}
\end{figure}
\par
Fig. \ref{fig:1} depicts the scalar power spectrum and $\fnl$ in the equilateral 
limit. 
One can see that for $k \leq k_{_{\rm LQC}}$, where $k_{_{\rm LQC}}$ is the 
scale set by the spacetime curvature at the bounce, the spectra are strongly scale dependent 
while for $k >> k_{_{\rm LQC}}$, the spectra approach their slow roll values. 
At low wave numbers, the figure shows that $\fnl$ is oscillatory. 
We have depicted the $\fnl$ for all configurations in Fig. \ref{fig:2}. 
In this figure, we have fixed the value of $k_1$ and varied $k_2$ and $k_3$ in 
such a way that they obey the triangle condition. 
This figure illustrates the shape of the non-Gaussianity. 
It can be seen that, in both the figures, the $\fnl$ peaks in the 
squeezed($k_3 \ll k_1\simeq\,k_2$) - flattened ($k_1\,\simeq\,k_2\,+\,k_3$) limit. 
\par
\begin{figure}
\begin{center}
\begin{tabular}{cc}
  \includegraphics[width=0.45\linewidth]{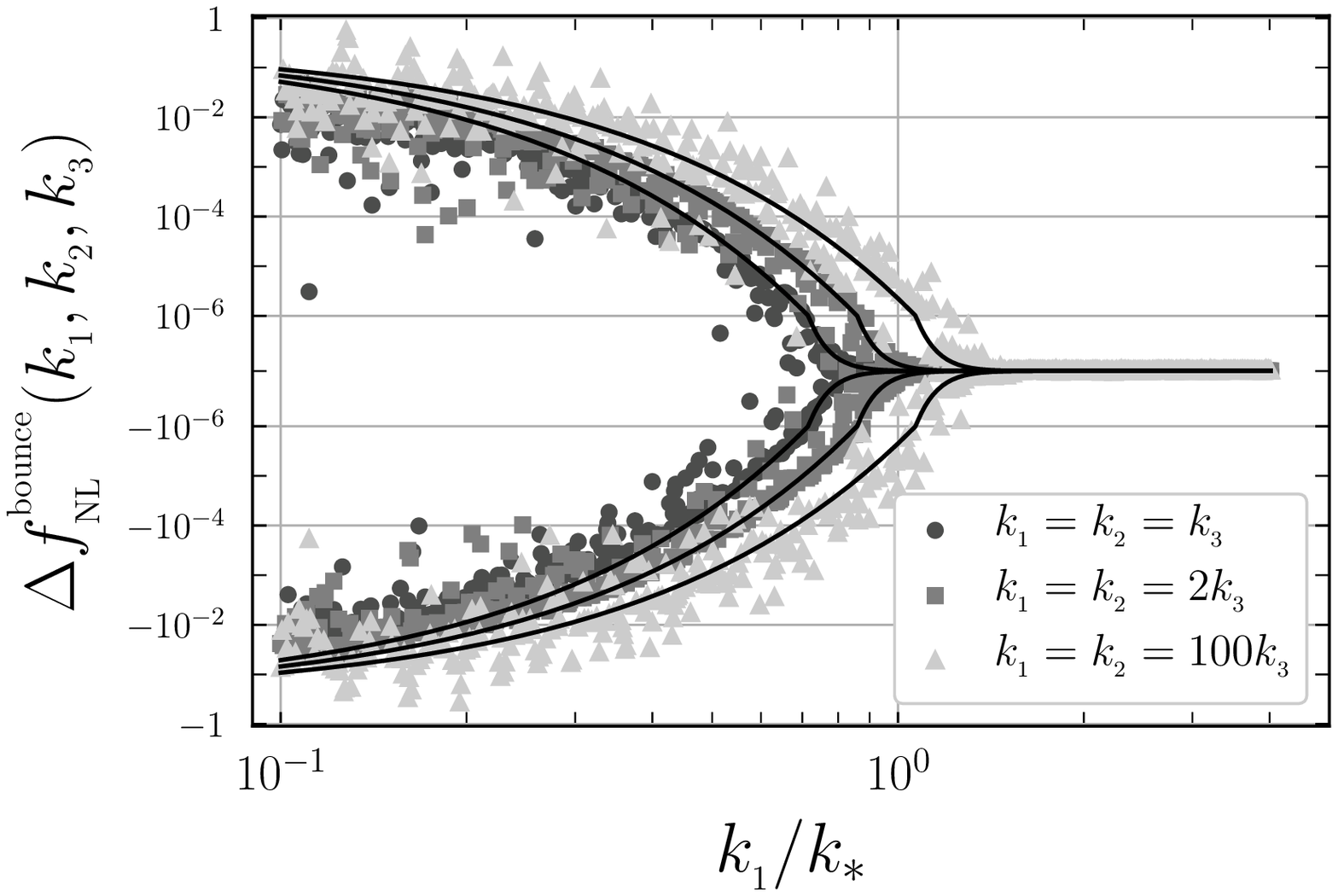} &
  \includegraphics[width=0.43\linewidth]{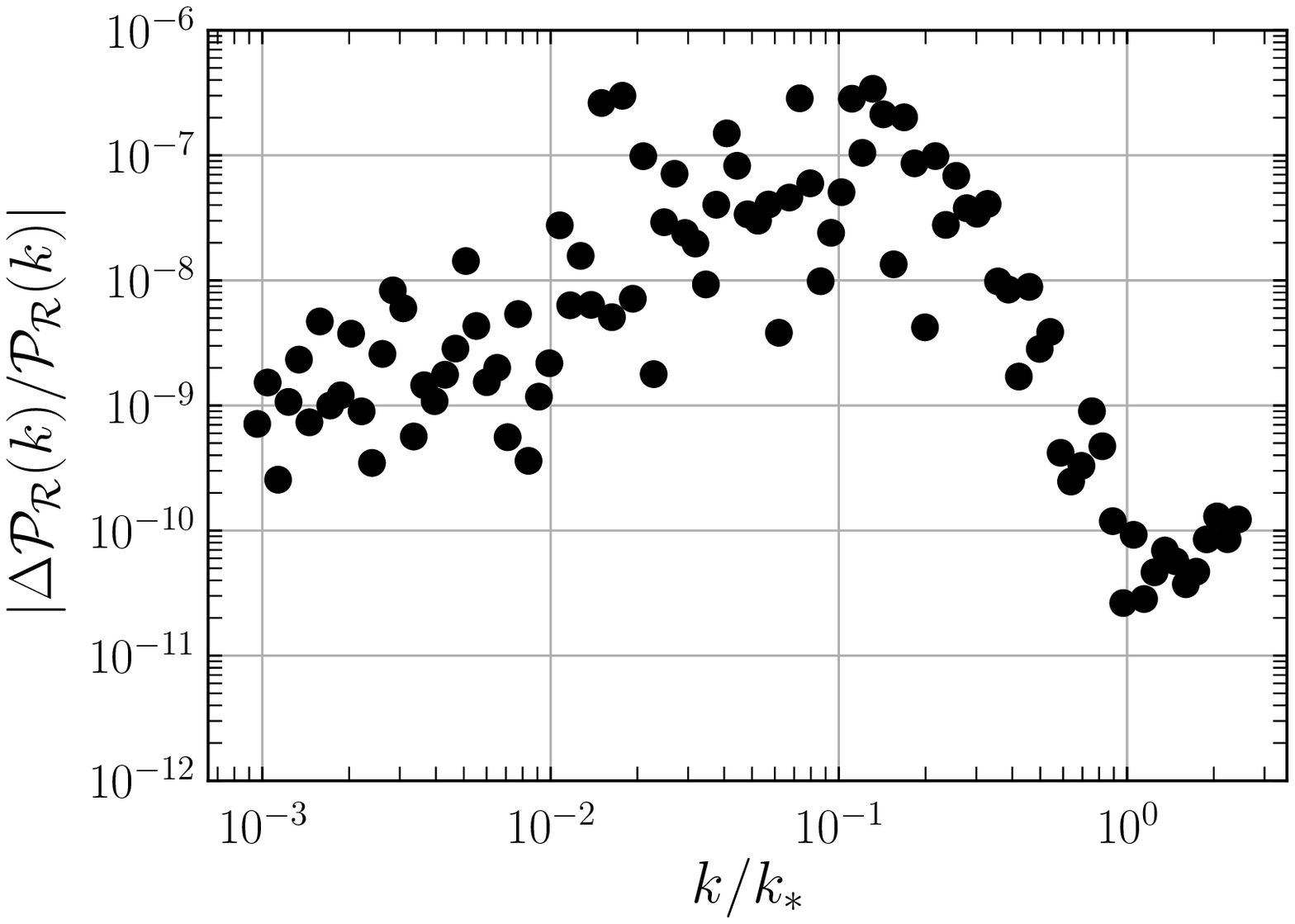}
  \end{tabular}
 \caption{\label{fig:3}On left, we have plotted a comparison of the analytical expression, $e^{- \alpha  k_t/k_{\rm LQC}}$, for 
 contribution to $\fnl$ from the bounce to the numerical result in different 
 configurations. 
 On right, we compute the relative amplitude of the leading order correction to the 
 power spectrum.}
 \end{center}
\end{figure}
The primordial non-Gaussianity generated due to LQC has a characteristic 
enhancement of amplitude at scales comparable to $k_{_{\rm LQC}}$. 
By analyzing the integrals involved in the computation of $\fnl$, we can estimate 
the contribution to $\fnl$ from the epoch around the bounce. 
For modes, $k \geq k_{_{\rm LQC}}$, we can approximate the mode function as 
$\varphi_k\,\sim\, e^{-ik\eta}$. 
Then the contribution to the integral from time around the  bounce can be schematically 
written as,
\begin{equation}
 \label{eqnr} I(k_1,k_2,k_3)\sim\int_{-\Delta \eta}^{\Delta \eta} \d\eta \, g(\eta)\, e^{i (k_1+k_2+k_3)\, \eta}
 \approx  \int_{-\infty}^{\infty} \d\eta \, g(\eta)\, e^{i k_t\, \eta} \, W(\eta,\Delta),
\end{equation}
 where, $g(\eta)$ is a combination of the functions depending on the background and the wavenumbers, 
 $k_t \equiv  k_1+k_2+k_3$ and $W(\eta,\Delta \eta)$ is a window function
 which selects only the contribution from the time range $-\Delta\eta<\eta<\Delta\eta$. 
 This integral can be computed using Cauchy's residue theorem and the 
 spectral dependence of the integral can be written as $e^{- \alpha  k_t/k_{\rm LQC}}$. 
 In Fig. \ref{fig:3}, we have compared the analytical expression with numerical result 
 and we find a good match between the two. 
\par
Finally let us compute the contribution to the power spectrum from the bispectrum. 
For the perturbation theory to be valid, this contribution has to be sub-dominant. 
The first perturbative correction to the two-point function of curvature perturbation 
is given by 
\begin{equation}
\langle 0|\h{\mathcal{R}}_{\vec k_1} \h{\mathcal{R}}_{\vec k_2}|0\rangle =
(2\pi)^3 \delta^{(3)}(\vec k_1+\vec k_2)\, \f{2\pi^2}{k_1^3} \, \hbar \,
\left[ \mathcal{P}_{\mathcal{R}}(k_1)+ \, \Delta \mathcal{P}_{\mathcal{R}}(k_1)\right] \, ,
\end{equation}
where
\begin{eqnarray}\label{eqn:DeltaP}
 \Delta \mathcal{P}_{\mathcal{R}}(k_1)&=&\hbar\, \f{k_1^3}{\pi^2}\, \Bigg[  \l(-\f{a}{z}\r)^3\, \left[-\f{3}{2}+3\f{V_{\phi}\, a^5}{\kappa\, \pp\, \pi_a}+\f{\kappa}{4}\f{z^2}{a^2}\right] \int\f{\d^3p}{(2\pi)^3}  \, B_{\delta\phi}(\vec{k}_1,\v p, -\v k_1-\v p)\,  \nonumber \\ 
 &+&  \l(-\f{a}{z}\r)^4\, \left[-\f{3}{2}+3\f{V_{\phi}\, a^5}{\kappa\, \pp\, \pi_a}+\f{\kappa}{4}\f{z^2}{a^2}\right]^2 \,   \int\f{\d^3p}{(2\pi)^3}\,  |\varphi_p|^2 \, |\varphi_{| \v k_1-\v p|} |^2\Bigg],
\end{eqnarray}
where $B_{\delta\phi}(\vec{k}_1,\v p, -\v k_1-\v p)$ is the bispectrum of inflaton perturbations and 
all the quantities on the right are evaluated at the end of inflation. 
We have numerically ploted the relative amplitude of the first order correction, 
$|\Delta \mathcal{P}_{\mathcal{R}}/\mathcal{P}_{\mathcal{R}}|$, in Fig. \ref{fig:3}. 
We find that, as expected, the magnitude of first-order correction to the power spectrum 
is negligible. 
This result can be qualitatively understood as follows. 
The leading order contribution to $\Delta \mathcal{P}_{\mathcal{R}}(k_1)$ is given by 
the first term in Eq. (\ref{eqn:DeltaP}) and it is given by $\epsilon\,\fnl\,\mathcal{P}_{\mathcal{R}}^2$, where 
$\epsilon$ is the slow roll parameter of $\mathcal{O}(10^{-2})$.
Since, $\fnl \leq 10^4$ and $\mathcal{P}_{\mathcal{R}} \leq 10^{-7}$, we obtain $\Delta \mathcal{P}_{\mathcal{R}}/\mathcal{P}_{\mathcal{R}} \leq 10^{-4}$ 
as in Fig. \ref{fig:3}.
\section{Discussion}\label{sec:4}
Let us conclude by making some remarks on the robustness of the results and 
its implication in the light of Planck data. 
We have verified the robustness of the results to a variation of the basic assumptions 
discussed in Sec. \ref{sec:3}.\cite{Agullo:2017eyh}
For instance, we find that the effect of changing $\phi_b$ is only a shift in the 
scale which is sensitive to the effect of the bounce with respect to the scales 
observable today. 
An increase in $\rho_B$ also leads only to a similar shift in the scales sensitive 
to the curvature of the bounce, in addition, to an increase in 
amplitude of $\fnl$. 
The Planck mission has put strong constraints on certain models of scale invariant 
non-Gaussianity, but, it provides little information on the scale dependent non-Gaussianity 
as produced in LQC.\cite{Ade:2015ava} 
Moreover, since the error bar on $\fnl$ goes as $1/\sqrt{\ell}$, at low multipoles, 
where the non-Gaussianity due to LQC is expected to be large, the error bar would be large. 
Considering the Planck error bars at low multipoles and demanding that 
the enhancement in $\fnl$ due to the LQC bounce appears at $\ell \lesssim 50$, 
one could try to arrive at constraints on the minimum value of scalar field at the bounce, 
for a given value of $\rhob$. 
Furthermore, by demanding that the imprint of the bounce should be at 
observable scales, we can arrive at an upper bound on $\phi_B$. 
For instance, for $\rhob\, =\, 1\, \Mpl^4$, we obtain $7.46 \Mpl \leq \phib \leq 7.82 \Mpl$. 
It should be kept in mind that the constraint described above is a very conservative estimate. 
Most probably, the oscillations in $\fnl$ will relax the constraint on $\phib$ discussed above. 
A more detailed account of this work has been published in Ref.~\refcite{Agullo:2017eyh}.
\section*{Acknowledgments}
We thank the organizers for giving VS the opportunity to present this work. 
VS would also like to thank Inter-University Centre for Astronomy and Astrophysics, Pune for financial 
support to attend MG XV. 
\bibliographystyle{ws-procs975x65}
\bibliography{main_refs}

\end{document}